\documentclass[%
 reprint,
 amsmath,amssymb,
 aps,
 prl,
 longbibliography,
 lengthcheck,%
]{revtex4-1}

\usepackage{graphicx}
\usepackage{dcolumn}
\usepackage{bm}
\usepackage{hyperref}

\begin{document}

\preprint{APS/123-QED}

\title{Thermodynamics: Extending and Reconstructing of Classical Theoretical Framework}%

\author{Tang Suye}
 \altaffiliation{}
 \email{tangsuye@hotmail.com}

\date{October 28, 2026}

\begin{abstract}
One deep problem that the traditional thermodynamics has faced since the beginning, I call it the predicament of
equilibrium state. In this predicament, both the initial and the final states of a
given system must be in equilibrium states, and the evolutionary processes from an initial
to an final states must be assumed to be a series of quasi-static processes, because in non-equilibrium
state, the temperature $T$ of a given system  and  the pressure $p$ of a given system are ill-defined, the real problem is, we did not know which extensive quantity that the temperature $T$ and the pressure $p$  are anchored to.

In the traditional thermo-dynamics: there has neither a state quantity of heat, nor dynamics.

This deep problem is derived from the absence of a fundamental physical quantity, the total energy of thermal motion in multiple levels, empirically be called
as the heat energy, as a state quantity, which has  never been defined on its own but only implied within the thermodynamic equations.
When the concept of the heat energy has been defined, the anchor that the temperature $T$ and the pressure $p$ are anchored to has been confirmed:
 the temperature $T$ and the pressure $p$ are the two intensive variables of the heat energy,
thus, the predicament of equilibrium state  has been completely broken through.

What we are seeing is a groundbreaking overall progress,  almost all the
long-standing tricky problems in the traditional thermodynamics will be finally clarified. In this article,
I will discuss some of the famous problems focusing on reconstructing a new theoretical framework: Thermodynamic processes are driven by the gradients of thermodynamic forces, the relationship between the states and the changes of
a thermodynamic system and the driving forces are the main themes of thermodynamics. What we see is a new narrative paradigm:  there have both the state quantity of the heat energy, and also dynamics.

\end{abstract}

\pacs{Valid PACS appear here}
\maketitle


\section{\label{sec:level1}Introduction\protect
}

One deep problem that the traditional thermodynamics has faced, I call it ``the predicament of equilibrium state''.

This predicament  is derived  from a fundamental issue: the intensive variables, the temperature $T$ and the pressure $p$ have
not been anchored to an interdependent extensive quantity, but been anchored to the thermodynamic system that is in
equilibrium state, namely ``the temperature $T$ of a given system '' and ``the pressure $p$ of a given system''.

An intensive variable denotes an intensity distribution of a given extensive quantity, for example, the elastic
stress  of a spring is an intensive variable that anchoring to the elastic potential energy of the spring, whereas,  in classical thermodynamics, we
don't know the intensity distributions that are indicated by the temperature $T$ and the pressure $p$  should be anchored to which extensive quantity.

In ``the predicament of equilibrium state'' of classical  theoretical framework, both the initial and the final states
of a given thermodynamic system must be in equilibrium states, and the evolutionary processes from an initial to a
final states must be assumed to be a series of quasi-static processes, because in non-equilibrium state, ``the temperature
of a system'' and ``the pressure of a system''  make no sense. In non-equilibrium thermodynamics, what we see is
an appeasement patching: assume local equilibrium, but this assumption cannot solve the deep problem that the two intensive variables,
the temperature $T$ and pressure $p$ have not been anchored to an interdependent extensive quantity.

The real problem is that in the traditional thermodynamic frameworks, we don't know which extensive quantity that
the temperature $T$ and the pressure $p$ are anchored to.

On the other hand, a process quantity must be anchored to a state quantity, this indicates an anchoring of the physical content, but in classical thermodynamics,
the process quantity $\delta Q$ has not been anchored to any state quantity.

In ``the predicament of equilibrium state'', both the initial and the final states of a given system must be in
equilibrium states, and the processes from any initial to any final states also must be a series of quasi-static processes,
this follows a restriction: we cannot  establish a dynamical  narrative to describe how a thermodynamic system evolves.

For example, for Fourier's law of heat conduction, the concept of the temperature gradient $\Delta T$ can hardly be compatible with
the assumptions of equilibrium state and quasi-static processes, even though the heat transfer should be one of the phenomena described by thermodynamics.

Friction generating heat is another example, an energy conversion process,  but no one can write an equation of the energy
conversion for the process, because the ``heat'' in here is not the heat exchange $\delta Q$. In classical thermodynamics,
there has no a concept that can describe heat production, due to there has no a state quantity of heat.

What we see is a complex  theoretical system which with a series of hard-to-understand problems and a series of appeasement mending that never been really made clear.

In over a century of the discussions on all of the famous problems, the deep problems on the  theoretical
framework  have never been addressed, what we see in the discussions are some derived problems,  never made real progress,
such as  anchoring the temperature $T$ and the pressure $p$ to an interdependent extensive quantity,
 establishing a new theoretical framework of dynamics,
introducing the concept of the entropy by a new approach to displace  heat engine cycles, confirming the thermodynamic description of
the physical content of the entropy and the second law, and, expounding the relationship between order production and the second law,  more importantly,
what we are seeing is such a thermo-dynamics which without dynamics.

The first thing we need to do is to make clear  how ``the predicament of equilibrium state" started:

1, We don't know which extensive quantity that the temperature $T$ and the pressure $p$ are anchored to.

2, Under the assumption of equilibrium state, the two intensive quantities, the temperature $T$ and the pressure $p$ are anchored to the  thermodynamic system,
become ``the temperature $T$ of a system'' and ``the pressure $p$ of a system''.

3, ``The temperature $T$ of a system'' and ``the pressure $p$ of a system'' are only applicable to equilibrium state,
so the evolutionary processes from an initial to a final states must be assumed to be a series of quasi-static processes.

4, The traditional thermodynamics is losing the way in ``the predicament of equilibrium state''.

5, And meanwhile, we don't know which state quantity that the process quantity $\delta Q$ is anchored to.

6, What we see is a very strange thermo-dynamics: there has neither a state quantity of heat, nor dynamics.

and more \ldots

The analysis above points to a clear direction, what we need to make clear first is whether the concept system of
extensive quantities in classical thermodynamics is already complete? then I realized, we don't know which extensive quantity
that temperature $T$ and pressure $p$ are anchored to,  and there has no a state quantity  for heat, the two are the same issue:
in classical thermodynamics, there is an indispensable extensive quantity that has not been defined.

This indispensable extensive quantity is exactly the total energy of thermal motion in multiple levels in a thermodynamic system, empirically be called as the heat energy.

It can easily be proved that the total energy of thermal motion in multiple levels within a given system is a state quantity,
the temperature $T$ and the pressure $p$  are the two intensive variables of the heat energy,
this seems to be a bit special because the most of the extensive quantities only have one intensive variable.

Such that, ``the predicament of equilibrium state'' of the traditional thermodynamics has been completely broken through.
\!Basing on the new progresses, combine with the known mechanical free energy and Gibbs free energy, we found
that the internal energy $U$ can be classified into the three independent forms, and  the equation of the first law can be switched
 to directly describe the states and the changes of the different forms of the internal energy,
 this means an important conclusion: the theoretical framework of the traditional thermodynamics has been reconstructed.

In a new theoretical framework, the narrative logic of thermodynamics has undergone significant changes,
what we are seeing is a dynamical narrative that is completely different from classical thermodynamics:
thermodynamic processes are driven by the gradients of thermodynamic forces, and the relationship between the state and changes of
a thermodynamic system and the driving forces are the main themes of thermodynamics.

The assumption of quasi-static processes has been completely discard, and equilibrium state is no longer thought to be a fundamental assumption,
 we have already broken through the predicament of  traditional  thermodynamics.

In a new theoretical framework, we extend the equations of the first law of thermodynamics, introduce the concept
of the  compensating potential, distinguish the energy transfer and the energy conversion processes by the concepts of the flux and the production,
we discuss the compensation relationships of the heat conversion, establish a complete narrative for the energy conversion processes.

In a new theoretical framework, we don't need to consider the assumption of equilibrium state and the assumption of the quasi-static processes,
because the temperature $T$ and the pressure $p$ have been confirmed to be the two intensive variables of the heat energy,
hence, we can discuss the energy transfer and the energy conversion from a new dynamical perspective, therefore, for examples,
 Fourier's law of heat conduction and Fick's law of diffusion both are the thermodynamic themes, this means a profound change, we can establish a theoretical framework of  dynamics.

In a new theoretical framework, the heat exchange $\delta Q$ has been anchored as the heat energy transfer between a system and its surroundings,
we can calculate the entropy changes caused by $\delta Q/T$ according to the state changes of the given system,
 as a derived concept, the definition of the entropy depend only on the equation of the first law, due to the  question that needs to be answered first is: what is it?
and according to the new definition, I proved that the entropy is the sum of the calorimetric entropy and the configurational entropy,
the physical meaning of the entropy is the sum of the two distribution functions, it can be proved in mathematics that the entropy is a state function, thus,
by the aid of the heat engine cycles to define the concept of the entropy\cite{Clausius1865} has been displaced by a new approach.

In a new theoretical framework, I have confirmed the total differential expression of the entropy production and the three sources of irreversibility.
Thermodynamic spontaneous processes are driven by different driving forces, the second law elucidates such a fundamental principle:
all of the gradients of the thermodynamic forces spontaneously tend to zero, in an irreversible process,
the dissipations of the gradients of the thermodynamic forces cannot
be spontaneous recovery completely, the total entropy production $d_iS \geq 0$ is the results of the irreversible processes.

In a new theoretical framework, we discuss the relationship between dissipation and evolution,
I found that the second law itself has already contained the mechanism of evolution, this point of view comes from such a fact,
in a coupling process,  we always have the total entropy production $d_iS\!\ge0$, but the partial differentials of $d_iS$ can be less than zero,
it implies that order and evolution can arise spontaneously, therefore, the second law is no longer thought to be only a dissipative law, it is also the law of evolution.

I believe, thermodynamics should be a self-consistent theory,  we are witnessing a profound change that will be brought into thermodynamics,
in which, all of the concepts and the principles are comprehensible.

``The fact that it is comprehensible is a miracle.''\cite{Einstein1936}

\section{Heat Energy: \,an implicit state quantity}
\label{}

In physics, thermal motion is a state property, and the energy of thermal motion is thought to be a form of energy, J.C.Maxwell\cite{Maxwell1872}, R.Feynman\cite{Feynmann1963} explicitly indicated that is the heat energy, and an earlier idea can also be found from D.Bernoulli book\cite{Bernoulli1738}: ``what we experience as heat is simply the kinetic energy of the great numbers of the particles motion'', as we know, this great numbers of the particles motion now is called as thermal motion.

However, in the traditional thermodynamics, the concept of the heat energy has never been defined on its own, it's only implied within the thermodynamic equations.

This is an fundamental concept in the whole theoretical framework of thermodynamics, if no a state definition on its own, thermodynamics will always face an unsolvable theoretical predicament, and, in fact, this predicament has being continued for more than a century.

Now, we consider how to define the energy of thermal motion on its own.

Given a system, using the symbol $q$ denotes the energy of thermal motion, named as the heat energy, which involves all of the heat phenomena at different levels.

By our experience, the heat energy is a non-conserved quantity, so it has the two sources, the energy transfer and the energy conversion, as we have already known, $\delta Q$ is the heat transfer, and the another source of the energy of thermal motion, we named as the heat production, comes from the heat conversion, such as a mechanocaloric effect, the heat effects of chemical reaction or phase transition.

Using $d_eq$ denotes the heat transfer, where the subscript $e$ denotes the energy transfer, we have $d_eq\,$=$\,\delta Q$,  using $d_iq$ denotes the heat production, we obtain
\begin{equation}
dq=d_eq+d_iq.
\end{equation}

In Eq.(1), the heat energy $q$ is a state quantity, the total changes of the heat energy of a given system are equal to the sum of the heat transfer and the heat production.

We consider an example first, for an ideal gas, all of the internal energy are the energy of thermal motion, therefore, in a reversible process, by the first law, we have
\begin{equation}
dU=dq=\delta Q-pdV.
\end{equation}
where $U$ is the internal energy, $\delta Q$ is the heat exchange, $p$ is the gas pressure, $V$ is the gas volume, and $-pdV\!=\!\delta W$ is the reversible volume work.

For an ideal gas,  the pressure $p$ comes from thermal motion, so $-pdV$ denotes the work done by the heat energy $q$, it signifies the heat energy of the gas to be converted into the work, and then exchanged with its surroundings.

For an ideal gas, $-pdV$ is an energy conversion process, the heat energy of the gas is converted into work via the $-pdV$ path, and the heat energy of the gas produces a tiny increment, that is the heat production $d_iq$, so we have
\begin{eqnarray}
d_eq=\delta Q=C_vdT.
\end{eqnarray}
\begin{eqnarray}
d_iq=-pdV.
\end{eqnarray}
where $C_v$ is the heat capacity at constant volume.

According to the state equation $pV$=$NkT$, the temperature $T$ and the pressure $p$ are the two intensive variables of the heat energy of an ideal gas, so the heat energy of an ideal gas can change via the two different paths.

Eq.(3) and Eq.(4) can be applied to the general case, that is, the temperature $T$ and the pressure $p$ are the two intensive variables of the heat energy $q$.

By our experience, the temperature $T$ denotes the intensity distribution of the energy of thermal motion, and the pressure $p$ involves density distribution of the energy of thermal motion, so the temperature $T$ and the pressure $p$  are the two intensive variables of the heat energy.

In general, the pressure denotes a force acting per unit area, which involves different sources, including thermal motion and the interactions between the particles. Using $p$ represents the heat partial pressure, $p$ comes from only the thermal motion, using $p^{\prime}$ represents the partial pressure when the other generalized force $Y$ is expressed as a force acting per unit area, the different sources of the total pressure therefore can be distinguished by
\begin{equation}
P=p+p^{\prime},
\end{equation}
where $P$ is the total pressure.

The first law thus becomes
\begin{eqnarray}
dU&=&\delta Q-pdV+Ydx+\sum_j\mu_{j}^{\prime}dN_j
\nonumber\\
&=&dq+Ydx+\sum_j\mu_{j}^{\prime}dN_j,
\end{eqnarray}
where $Y$ is the generalized force, $dx$ is the generalized displacement, $Ydx$ includes $-p'dV$, $\mu_{j}^{\prime}$ is the chemical potential of the type\,-\emph{j} particles, and $N_j$ is the number of such particles, corresponds to $\mu_{j}^{\prime}\neq 0$. The internal energy $U$ is the sum of the three independent forms of the energies.

An interesting difference shows that ``the pressure $p$ has different sign from other generalized force, if we increase the pressure, the volume increases, whereas if we increase the force, \!$Y$, for all other cases, the extensive variable, $x$, decreases''\cite{Reichl1997}. This is an interesting criterion that can be used to distinguish the heat partial pressure $p$ from other generalized force $Y$, by this criterion, the heat partial pressure $p$ will be different from other generalized force.

We find that $dq$ can be proven to be an exact differential.

Consider $U$=$U(T, V, x,N_j)$, where $T,V,x$ and $N_j$ are the four independent state variables. Then we obtain\cite{Reichl1997}
\begin{eqnarray}
dU&=&\!\left(\frac{\partial U}{\partial T}\right)_{V,x,N_j}\!\!\!dT+\!\left(\frac{\partial U}{\partial V}\right)_{T,x,N_j}\!\!\!dV
    \nonumber\\
&&\!+\!\left(\frac{\partial U}{\partial x}\right)_{T,V,N_j}\!\!\!dx+\!\sum_j\left(\frac{\partial U}{\partial N_j}\right)_{T,V,x,N_{i\neq j}}\!\!\!\!dN_j.\,\,\,\,\,\,\,\,\,
\end{eqnarray}

By definition and compare Eq.(7) with Eq.(6) , $q$ is the function of the two independent variables $T$ and $V$
\begin{eqnarray}
dq=\left(\frac{\partial q}{\partial T}\right)_{V}dT+\left(\frac{\partial q}{\partial V}\right)_{T}dV.
\end{eqnarray}

Compare Eq.(7) with Eq.(8), we get
\begin{eqnarray}
dq&=&\left(\frac{\partial q}{\partial T}\right)_{V}dT+\left(\frac{\partial q}{\partial V}\right)_{T}dV
        \nonumber\\
&=&\left(\frac{\partial U}{\partial T}\right)_{V,x,N_j}\!\!\!dT+\left(\frac{\partial U}{\partial V}\right)_{T,x,N_j}\!\!\!dV.
\end{eqnarray}

Owing to the fact that $dU$ is an exact differential, by Eq.(7) and Eq.(9), the two partial derivatives of the internal energy satisfy the relation that
\begin{eqnarray}
\left[\frac{\partial}{\partial V}\left(\frac{\partial U}{\partial T}\right)_{V,x,N_j}\right]_{T,x,N_j}\!
=\!\left[\frac{\partial}{\partial T}\left(\frac{\partial U}{\partial V}\right)_{T,x,N_j}\right]_{V,x,N_j}
\end{eqnarray}

Such that, we have
\begin{eqnarray}
\left[\frac{\partial}{\partial V}\left(\frac{\partial q}{\partial T}\right)_V\right]_{T}
=\left[\frac{\partial}{\partial T}\left(\frac{\partial q}{\partial V}\right)_T\right]_V.
\end{eqnarray}

Thus we have proven that $dq$ is an exact differential.

Using the function $A$ denotes the mechanical free energy within a given system that
\begin{equation}
A=Yx.
\end{equation}

The equation of the first law becomes
\begin{eqnarray}
dU&=&dq+dA+dG
\nonumber\\
&=&dq+d(Yx)+\sum_jd(\mu_{j}^{\prime} N_j).
\end{eqnarray}

The internal energy $U$ is equal to the sum of the internal heat energy $q$, the internal mechanical free energy $A$ and the internal Gibbs free energy $G$.

In Eq.(13), since $Y$ and $x$ both are the two independent variables, we have
\begin{eqnarray}
\left[\frac{\partial}{\partial Y}\left(\frac{\partial A}{\partial x}\right)_Y\right]_{x}
=\left[\frac{\partial}{\partial x}\left(\frac{\partial A}{\partial Y}\right)_x\right]_Y
=1.
\end{eqnarray}

So, the internal mechanical free energy $A$ is a state function, and similar for Gibbs free energy $G$.

$A$ and $G$  both are the non-conserved quantities, so we have the two equations similar to Eq.(1)
\begin{eqnarray}
dA=d_eA+d_iA.
\end{eqnarray}
where $d_eA$ denotes the mechanical free energy flux, and $d_iA$ denotes the mechanical free energy production.

For Gibbs free energy we have
\begin{eqnarray}
dG=d_eG+d_iG.
\end{eqnarray}
where $d_eG$ denotes Gibbs free energy flux, $d_iG$ denotes Gibbs free energy production, similar to Eq.(15).

Such that, the concepts ``flux'' and ``production'' can be used as the fundamental  concepts to distinguish the two processes: the energy transfer and the energy conversion.

The equation of the first law can be extended that
\begin{eqnarray}
dU=dq+dA+dG.
\end{eqnarray}
\begin{eqnarray}
dU=d_eq+d_eA+d_eG.
\end{eqnarray}
\begin{eqnarray}
d_iq+d_iA+d_iG=0.
\end{eqnarray}

The three equations are the purely extensive quantities equations,  such that, the predicament of equilibrium state of the traditional thermodynamics has been  broken through completely start from the equations of the first law.

We have already known that the temperature $T$ and the pressure $p$ are the two  intensive variables of the heat energy, so the two fundamental assumptions of classical thermodynamics, the assumptions of equilibrium state and quasi-static processes are therefore totally meaningless.

According to Eq.(19), we have the following relation
\begin{eqnarray}
d_iq=-(d_iA+d_iG).
\end{eqnarray}

The heat production is equal to the negative free energy productions due to conservation of energy.

As an extension, we introduce the concept of the free energy to distinguish the state properties of different forms of thermodynamic energies,
the criterion is whether a given form of the thermodynamic energy has a quality loss, in the later discussions,
a further criterion is whether the entropy of the given form of the thermodynamic energy is zero.

According to this criterion, the mechanical free energy $A$ and Gibbs free energy $G$ are the free energies, and the heat energy $q$ is the non-free energy.

Since the non-free energy has has a quality loss, which needs compensation when converting into the free energy.

The driving force of the heat energy converting into the free energy, such as heat into work,
comes from the heat partial pressure $p$, there are two changes within the system, one is the heat production $d_iq$, and another one is $pdV$.

In a reversible conversion process, we have
\begin{eqnarray}
pdV=-d_iq.
\end{eqnarray}

However, in an irreversible process, we have
\begin{eqnarray}
pdV\geq-d_iq.
\end{eqnarray}

This is the result with some spontaneous dissipation.

Since the heat energy $q$ has the two intensity variables $T$ and $p$, only one can be used in the energy conservation equation, otherwise $q$ or $dq$ would be calculated twice.

In a reversible process, $-pdV$ has the two meanings: $-pdV$ itself is the volume work; in Eq.(21), it represents a compensation for an energy conversion process.

I named $pdV$ as the change of a compensating potential,
the reversible part provides a conversion compensation for the non-free energy converting into the free energy, Eq.(21) is a compensation relation,
in Eq.(22), the irreversible part itself has no the volume work done but only indicates the spontaneous dissipation of the compensating potential.

\section{Thermo-dynamics: an Overview}

In the traditional thermodynamics, no one can establish a theoretical framework of dynamics in the predicament  of equilibrium state,
so  what we see is such a situation, there has neither a state quantity of heat, nor dynamics.

Under the assumptions of equilibrium state and quasi-static processes, thermodynamic driving forces, such as the temperature gradient $\Delta T$ and the pressure gradient $\Delta p$ are ill- defined. we can only calculate the difference  between the final equilibrium state and the initial equilibrium state, but we can not start from an initial state and other parameters to deduce how a given system will change.

Thermo-dynamics should be such a physics theoretical system at macroscopic level, which describes the relations between the thermodynamic driving forces and the state changes for a multi-body thermodynamic system.

In the view of dynamics, the main themes of thermo-dynamics are to describe the states and changes of a given thermodynamic system, including following subjects:

1, The energy transfer and diffusion processes.

2, The mass transfer and diffusion processes.

3, The energy conversion processes.

4, The generating of the structure and the configuration of matter at different levels.

Now let's talk about the first three subjects, the fourth subject is too magnificence.

We start from the energy transfer or diffusion processes, the driving forces of thermodynamics are the temperature gradient $\nabla T$ and the pressure gradient $\nabla p$.

The heat transfer process driven by $\nabla T$ has already been given by Fourier's law of heat conduction\cite{Fourier1878}.
\begin{eqnarray}
\emph{\textbf{J}}_q=\lambda_T \nabla T.
\end{eqnarray}
where $J_q$ is the heat flux density that flows through per unit cross-sectional area at per unit time, the driving force is the temperature gradient  $\nabla T$,  $\lambda_T$ is the thermal conductivity.

$\nabla T$ denotes an intensity gradient of the heat energy, or a macroscopic energy levels gradient.

The subscript $T$ denotes that the heat transfer is driven by the temperature gradient.

The gradient of a thermodynamic force is a vector, which applies to the specific directional conventions of physical quantities in physics. The direction of the force vector is along the line of action pointing towards the point or object being acted on, which is different from the mathematical convention of the gradient direction, so in Eq.(23), there is no minus sign ``-" on the right-hand side, which is different from the mathematical convention but more conciseness.

Eq.(23) is only a different way of expressing Fourier's law of heat conduction, the physical meanings are same.

Since the heat energy has another intensive variable $p$, the pressure gradient $\nabla p$ is in relation to a density gradient of the heat energy, so the pressure gradient $\nabla p$ is also the driving force of the heat transfer, we have
\begin{eqnarray}
\textbf{\emph{J}}_q=\lambda_p \nabla p.
\end{eqnarray}
where $J_q$ is the heat flux density that flows through per unit cross-sectional area at per unit time, the driving force is the pressure gradient $\nabla p$, and $\lambda_p$ is the coefficient of thermal diffusion.

The subscript $p$ denotes that the heat transfer is driven by the pressure gradient.

This means that there are two driving forces of the heat diffusion phenomena, so the distribution and changes of the heat field are related to both the temperature field $\nabla T$ and the pressure field $\nabla p$ at the same time.

The heat flux flowing into a given volume element
\begin{equation}
\emph{\textbf{J}}_q= \lambda_T\nabla T+ \lambda_p\nabla p.
\end{equation}

At the microscopic level, atoms, molecules, and other particles are the carriers of the heat energy, so that there is also Fick's law of diffusion
\begin{equation}
\textbf{\emph{J}}=D\nabla C.
\end{equation}
where, $J$ is the diffusion flux per unit cross-sectional area at per unit time, $D$ is the diffusion coefficient, $\nabla C$ is the gradient of the particle number density or concentration. Here,  according to  the specific directional conventions of special vector in physics, there is no negative sign.

Eq.(26) can also be written as
\begin{equation}
\emph{\textbf{J}}_m= \lambda_{Tm}\nabla T+ \lambda_{pm}\nabla p.
\end{equation}

The subscript $m$ denotes the mass transfer or diffusion.

The subscripts $T$ and $p$ have the same meanings as that in Eq.(23) and Eq.(24).

For a mixed system, it can be calculated separately based on the particle type and then sum up, such as a mixed gas.

Now let's talk about the third subject, the energy conversion processes.

By Eq.(20), we have
\begin{eqnarray}
d_iq=-(d_iA+d_iG).
\end{eqnarray}

Eq.(28) shows that the heat production is equal to the negative free energy productions.

There are two situations in the energy conversion processes: (1), the free energies $A$ or $G$ convert into the heat energy $q$, (2), the heat energy $q$ is converted into the free energies $A$ or $G$ driving by the compensating potential. The thermodynamic driving forces of the two conversion processes are different.

In the first situation, the free energies $A$ and/or $G$ convert into the heat energy $q$,  the thermodynamic driving forces of the
conversion processes are the  thermodynamic gradients  $\Delta Y$ and/or $\Delta \mu^{\prime}$, we have
\begin{eqnarray}
d_iq=\Delta Ydx+\sum_j\Delta \mu_{j}^{\prime} dN_j.
\end{eqnarray}

For example, regarding the mechanical thermal effects, such as Friction generating heat, we have
\begin{eqnarray}
d_iq=-d_iA=\Delta Ydx.
\end{eqnarray}

For the heat effects of the chemical reactions or phase changes, we get
\begin{eqnarray}
d_iq=-d_iG=\sum_j\Delta \mu_{j}^{\prime} dN_j.
\end{eqnarray}

In the second situation,  the heat energy $q$ is converted into the free energies $A$ and/or $G$, in these situations, what we see is a coupling process, a coupling process means that there are two or more gradients of thermodynamic forces participating in the processes.

When the heat energy $q$ is converted into the free energies $A$ and/or $G$, the equation of the energy conversion is the same as Eq.(29)
\begin{eqnarray}
d_iq=\Delta Ydx+\sum_j\Delta \mu_{j}^{\prime} dN_j.
\end{eqnarray}

In this situation, the heat production $d_iq\leq 0$, and the free energies productions $d_iA+d_iG\geq 0$,
the gradients $\Delta Y$ and/or $\Delta \mu_{j}^{\prime}$ are increasing, so the two both are not the driving forces of the conversion processes, which need another thermodynamic gradient to push the processes.

In the second situation, the driving force is the pressure gradient $\Delta p$, here we discuss only the energy conservation equation, the compensation relation is shown in Eq.(21).

Combine with the energy transfer equations, we obtain
\begin{eqnarray}
dA=d_eA+d_iA=Ydx-\Delta Ydx.
\end{eqnarray}
\begin{eqnarray}
dG=d_eG+d_iG=\sum_j\mu_{j}^{\prime} dN_j-\sum_j\Delta \mu_{j}^{\prime} dN_j.\,\,\,
\end{eqnarray}
where, $d_eA$ and $d_eG$ both are the free energy transfer, $d_iA$ and $d_iG$ both are the free energy production.

When $d_iq\geq0$, $d_iA$ and $d_iG$ can be used to calculate the dissipations in the free energies transfer processes.

More in-depth topics, such as the compensation relation of the energy conversion, the spontaneous dissipation of the compensating potential,
especially for the mechanism of order production, will be elucidated in more detail when discussing on the total entropy production and the internal mechanism of the second law of thermodynamics.

\section{\label{sec:level1}Entropy: Exact Differential\protect
}

In classical thermodynamics, the concept of the entropy was defined by the aid of Carnot cycle\cite{Clausius1865} that
\begin{eqnarray}
\oint_{rev} \frac{\delta Q}{T}=0.
\end{eqnarray}
where $rev$ denotes the reversible processes. It given a conditional definition with an additional restriction.
\begin{eqnarray}
dS=\bigg(\frac{\delta Q}{T}\bigg)_{rev}.
\end{eqnarray}

In  Eq.(36), the definition $dS\!=\!\delta Q/T_{rev}$ is the reversible path dependent, it should be noted that $\delta Q/T$ is a process quantity, and is not  anchored to any state quantity.

We now consider a new approach.

Consider the heat transfer $\delta Q$, it now is known as the change of an extensive quantity $q$, and can be transferred spontaneously only from higher temperature to lower temperature, this property shows us that the heat transfer $\delta Q$ relates to the intensity distribution of the temperature $T$.

For the same $\delta Q$, if its temperature distribution is different, the intensive property of $\delta Q$ will be also different, thus, we can use the function $\delta Q/T$ to measure $\delta Q$ and its intensive property of the temperature distribution, then $\delta Q/T$ may be named as the entropy of the heat flux $\delta Q$.
\begin{equation}
d_eS=\frac{\delta Q}{T}.
\end{equation}

In Eq.(37), $\delta Q/T$ is only a process quantity. For a given system, we need to consider when $\delta Q$ to be transformed into the internal energy of a given system, or the internal energy of the system to be transformed into the heat flux $\delta Q$, how do we describe the state changes of the system.

By Eq.(6), it follows that
\begin{equation}
\frac{\delta Q}{T}\Rightarrow \frac{dU}{T}-\frac{Ydx}{T}-\sum_j \frac{\mu_{j}^{\prime} dN_j}{T}+\frac{pdV}{T}
 \nonumber\\
\Rightarrow \frac{dq}{T}+\frac{pdV}{T}.
\end{equation}

When $\delta Q/T$ to be transferred into the given system, the right hand side of the equation is the state changes.

Such that, we can define a state function according to the state changes of the given system.

Using a function  denotes the state changes
\begin{equation}
dS=\frac{dq}{T}+\frac{pdV}{T}.
\end{equation}

$S$ is named as the entropy of the given system.

Eq.(38) is a binomial expression, the integral of $dq/T$ we named as the calorimetric entropy $S_q$.
\begin{equation}
dS_q=\frac{dq}{T}.
\end{equation}

The integral of $pdV/T$ we named as the configurational entropy  $S_v$, the configurational entropy $S_v$ is in relation to the density distribution of the particles number or the density distribution of the heat energy.
\begin{equation}
dS_v=\frac{pdV}{T}.
\end{equation}

By Eq.(8) and Eq.(38), $S$ is the function of the two independent variables $T$ and $V$.
According to the mathematical theory of exact differential, if the two partial derivatives of the function $S$  satisfy the equation
\begin{equation}
\left[\frac{\partial}{\partial V}\left(\frac{\partial S}{\partial T}\right)_{V}\right]_{T}=\left[\frac{\partial}{\partial T}\left(\frac{\partial S}{\partial V}\right)_{T}\right]_{V}.
\end{equation}

$dS$ will be an exact differential.

We now prove this conclusion.

Using Eq.(8) in Eq.(38), it follows that
\begin{eqnarray}
dS&=&\frac{dq}{T}+\frac{p}{T}dV
 \nonumber\\
&=&\frac{1}{T}\left(\frac{\partial q}{\partial T}\right)_VdT+\frac{1}{T}\left(\frac{\partial q}{\partial V}\right)_TdV+\frac{p}{T}dV.\,\,\,\,\,
\end{eqnarray}

By Eq.(42) we have
\begin{eqnarray}
\left[\frac{\partial}{\partial V}\left(\frac{\partial S}{\partial T}\right)_{V}\right]_{T}&=&\left\{\frac{\partial}{\partial V}\left[\frac{1}{T}\left(\frac{\partial q}{\partial T}\right)_V\right]\right\}_{T}
 \nonumber\\
&=&\frac{1}{T}\left[\frac{\partial}{\partial V}\left(\frac{\partial q}{\partial T}\right)_V\right]_{T}.
\end{eqnarray}

and
\begin{eqnarray}
\!\left[\frac{\partial}{\partial T}\left(\frac{\partial S}{\partial V}\right)_{T}\right]_{V}\!\!&=&\!
\left\{\frac{\partial}{\partial T}\left[\frac{1}{T}\left(\frac{\partial q}{\partial V}\right)_T+\left(\frac{p}{T}\right)\right]\right\}_V
                 \nonumber\\
&=&\!\frac{1}{T}\left[\frac{\partial}{\partial T}\left(\frac{\partial q}{\partial V}\right)_T\right]_V-\frac{1}{T^2}\left(\frac{\partial q}{\partial V}\right)_T
                 \nonumber\\
&&+\frac{1}{T^2}\left[T\left(\frac{\partial p}{\partial T}\right)_V-p\right].
\vspace{0.1cm}
\end{eqnarray}

Before we compare Eq.(43) with Eq.(44), we need to prove a similar relation in classical thermodynamics anew
\begin{equation}
\left(\frac{\partial q}{\partial V}\right)_{T}=T\left(\frac{\partial p}{\partial T}\right)_{V}-p.
\end{equation}

From Eq.(42) we have
\begin{eqnarray}
dS=\frac{1}{T}\left(\frac{\partial q}{\partial T}\right)_VdT+\frac{1}{T}\left[\left(\frac{\partial q}{\partial V}\right)_T+p\right]dV.
\end{eqnarray}

It follows that
\begin{eqnarray}
dS=\left(\frac{\partial S}{\partial T}\right)_VdT+\left(\frac{\partial S}{\partial V}\right)_TdV.
\end{eqnarray}

$dq$ can be expressed as the follows
\begin{eqnarray}
dq=TdS-pdV.
\end{eqnarray}

The pressure $p$ and the volume $V$  both are the state variables, so $d(pV)$ is an exact differential, we consider
\begin{eqnarray}
pV=TS-q.
\end{eqnarray}
\vspace{-0.6cm}
\begin{eqnarray}
d(pV)=d(TS)-dq=TdS+SdT-dq.
\end{eqnarray}

Using Eq.(48) in Eq.(50), we get
\begin{eqnarray}
d(pV)=d(TS)-dq=SdT+pdV.
\end{eqnarray}

Where $d(pV)$=$d(pV)(T,V)$, we have from Eq.(51)
\begin{eqnarray}
\left[\frac{\partial }{\partial V}\left(\frac{\partial (pV)}{\partial T}\right)_V\right]_T=\left(\frac{\partial S}{\partial V}\right)_T.
\end{eqnarray}

and
\begin{eqnarray}
\left[\frac{\partial }{\partial T}\left(\frac{\partial (pV)}{\partial V}\right)_T\right]_V=\left(\frac{\partial p}{\partial T}\right)_V.
\end{eqnarray}

Since $d(pV)$ is an exact differential, so that
\begin{eqnarray}
\left(\frac{\partial S}{\partial V}\right)_T=\left(\frac{\partial p}{\partial T}\right)_V.
\end{eqnarray}

Using Eq.(54) in Eq.(47), we get
\begin{eqnarray}
dS=\left(\frac{\partial S}{\partial T}\right)_VdT+\left(\frac{\partial p}{\partial T}\right)_VdV.
\end{eqnarray}

Compare Eq.(55) with Eq.(46), the two partial derivatives satisfy the equation that
\begin{eqnarray}
\left(\frac{\partial p}{\partial T}\right)_V=\frac{1}{T}\left[\left(\frac{\partial q}{\partial V}\right)_T+p\right].
\end{eqnarray}

It follows that
\begin{eqnarray}
\left(\frac{\partial q}{\partial V}\right)_T=T\left(\frac{\partial p}{\partial T}\right)_V-p.
\end{eqnarray}

Thus we have proven the relation Eq.(45).

Using Eq.(57) in Eq.(44), we get
\begin{eqnarray}
\left[\frac{\partial}{\partial T}\left(\frac{\partial S}{\partial V}\right)_{T}\right]_{V}
=\frac{1}{T}\left[\frac{\partial}{\partial T}\left(\frac{\partial q}{\partial V}\right)_T\right]_V.
\end{eqnarray}

Compare Eq.(43) and Eq.(58) with Eq.(11), we obtain the following conclusion
\begin{equation}
\left[\frac{\partial}{\partial V}\left(\frac{\partial S}{\partial T}\right)_V\right]_{T}=\left[\frac{\partial}{\partial T}\left(\frac{\partial S}{\partial V}\right)_{T}\right]_V.
\end{equation}

Eq.(59) indicates that $dS$ is an exact differential, $S$ is a state function, the conclusion perfectly obey the general rules of the exact differential, and does not need to depend on an imaginary reversible cycle, such that, the traditional approach has been displaced by a new approach.

As the conclusion of Eq.(59), the function changes $dS$ will be path-independent, so the integral of $dS$ around an arbitrary closed path will be always equal to zero!
\begin{eqnarray}
\oint dS\equiv0.
\end{eqnarray}

We now consider some examples. The first example is a photon gas. Using $q$=$\alpha T^4V$, $p$=$\alpha T^4/3$ in Eq.(32), where $\alpha$ is a constant, and then we get that
\begin{eqnarray}
dq=d(\alpha T^4V)=4\alpha T^3VdT+\alpha T^4dV.
\end{eqnarray}
\vspace{-0.5cm}
\begin{eqnarray}
dS&=&\frac{1}{T}dq+\frac{p}{T}dV
  \nonumber\\
&=&\frac{1}{T}4\alpha T^3VdT+\frac{1}{T}\alpha T^4dV+\frac{1}{3T}\alpha T^4dV
 \nonumber\\
&=&4\alpha T^2VdT+\frac{4}{3}\alpha T^3dV.
\end{eqnarray}

We have from Eq.(62)
\begin{equation}
\left[\frac{\partial}{\partial V}\left(\frac{\partial S}{\partial T}\right)_V\right]_{T}=\left[\frac{\partial}{\partial T}\left(\frac{\partial S}{\partial V}\right)_{T}\right]_V=4\alpha T^2.
\vspace{0.01cm}
\end{equation}

Thus the entropy of a photon gas is a state function.

The second example is an ideal gas. \!Using $dq$=$C_vdT$\,\, and $pV$=$nRT$ in Eq.(42), where $ C_v$ is the heat capacity at constant volume, $n$ is the number of moles, and $R$ is the gas constant. We have
\begin{eqnarray}
dS&=&\frac{C_v}{T}dT+\frac{p}{T}dV
  \nonumber\\
&=&\frac{C_v}{T}dT+\frac{nR}{V}dV.
\end{eqnarray}

and
\begin{equation}
\left[\frac{\partial}{\partial V}\left(\frac{C_v}{T}\right)\right]_{T}=\left[\frac{\partial}{\partial T}\left(\frac{nR}{V}\right)\right]_V=0.
\vspace{0.01cm}
\end{equation}

Thus the entropy of an ideal gas is a state function.

For an ideal gas, $T$=$2q/iNk$, $pV$=$NkT$, we get
\begin{eqnarray}
dS=\frac{i}{2}\frac{Nk}{q}dq+\frac{Nk}{V}dV.
\end{eqnarray}
where $i$ is the number of the degrees of freedom of the gas molecules, $N$ is the number of the gas molecules, and $k$ is Boltzmann constant.

For a photon gas, we have the photon gas equation
\vspace{-0.1cm}
\begin{eqnarray}
pV=\frac{\zeta (4)}{\zeta (3)}NkT=NR_{\{p\}}T.
\vspace{-0.1cm}
\end{eqnarray}
where $N$ is the number of the photons, $\zeta (3)$ and $\zeta (4)$ both are the Riemann zeta functions,  $R_{\{p\}}$=$[\zeta (4)/\zeta (3)]k$ is the photon gas constant.

Consider $q$=$3pV$ and $T$=$q/3NR_{\{p\}}$, combine Eq.(67) with Eq.(62), we get
\begin{eqnarray}
dS=\frac{3NR_{\{p\}}}{q}dq+\frac{NR_{\{p\}}}{V}dV.
\end{eqnarray}

All of the variables in Eq.(66) and Eq.(68) are extensive variables, where $q/iN$ refers to  the average energy levels per degree of freedom of the particles,
$N/V$ is the density function of the particles. In an earlier work\cite{Suye2008}, I presented a mathematical model to express the average degrees of the distribution of the two extensive quantities $q$ and $N$, for an ideal gas or a photon gas, the Nature of the Entropy is a binomial distributions function.

According to Eq.(66) and Eq.(68), the explanation of the binomial distribution function can easily be extended to the microscopic level, this means that there are further microscopic explanations, thermodynamics itself can explain the physical meaning of the concept of the entropy.

In Eq.(66) and Eq.(68), $dS$ has the same function form, it may be the primary equation of the entropy that
\begin{eqnarray}
dS=\frac{\hat{\alpha}NR_{\{k\}}}{q}dq+\frac{NR_{\{k\}}}{V}dV.
\vspace{-0.2cm}
\end{eqnarray}
where $\hat{\alpha}$ denotes a factor, for an ideal gas $\hat{\alpha}\!=\!i/2$, for a photon gas $\hat{\alpha}\!=\!i$, and $R_{\{k\}}$ denotes the system constant, for an ideal gas $R_{\{k\}}$=$k$, for a photon gas $R_{\{k\}}$=$R_{\{p\}}$.

According to Eq.(59), Eq.(38) is a perfect definition of the state function in the usual sense.

In traditional thermodynamics, we can not explain the physical contents of the entropy by thermodynamics itself, because $\delta Q/T$ has not been anchored to a state quantity.

This used to be a century-old mystery, I.Prigogine once wrote in 1989 : ``Now entropy is a very strange concept and without hoping to achieve a complete description''\cite{Prigogine1989}.

Now, we will be able to solve this century mystery.

Thermodynamic entropy is the sum of a binomial distribution function: one is the function of the energy levels distribution of the heat energy $q$,
 another one involves the distribution function related to the heat energy density or the particle number density,
 and the value of the entropy is proportional to the average degrees of the two distributions.

Where, the macro energy level is the temperature $T$.

\section{\label{sec:level1}The second law: The Total Differential\protect
}
The classical inequality of the second law is derived by the aid of heat engine cycle, it is expressed as the increment in the entropy of an isolated system $dS\!\geq\!0$, and in non-equilibrium thermodynamics, the increment in the entropy of a given isolated system has been displaced by the total entropy production $d_iS$ and
the inequality $d_iS\!\geq\!0$.

In classical thermodynamics, Clausius inequality dose not have an  explicit total differential,  due to $\delta Q/T$ has not been anchored to
a state quantity, and in non-equilibrium thermodynamics, we still don't know which state quantities that the entropy $S$ are anchored to. In the absence of the independent definition of the heat energy,  the evolutionary mechanism of the second law is unclear.

Let's first consider how to determine the total differential of the total entropy production.

By Eq.(38) and Eq.(1), we have
\begin{equation}
dS=\frac{dq}{T}+\frac{pdV}{T}=\frac{d_eq}{T}+\frac{d_iq}{T}+\frac{pdV}{T}.
\end{equation}

The total entropy production is a process quantity, we consider a tiny process, suppose the given system changes from the state $A$ to the state $B$, we have
\begin{eqnarray}
d_iS\!=\!\!\bigg(\frac{1}{T_B}\!-\!\frac{1}{T_A}\bigg)d_eq\!+\!\frac{d_iq}{T}\!+\!\!\bigg(\frac{p_B}{T_B}-\!\frac{p_A}{T_A}\bigg)dV\!.\;\;\;\;\;
\end{eqnarray}
where, the heat flux $d_eq$ is a micro-quantity $dq$ in transfer, can be written as $dq$,
the third term in the equation we consider $p/T$ as an integrand, and $V$ is an integral variable.

By Eq.(29) and Eq.(32), we have
\begin{eqnarray}
d_iq=\Delta Ydx+\sum_j\Delta \mu_{j}^{\prime} dN_j.
\end{eqnarray}

Using Eq.(72) in Eq.(71),  it follows that
\begin{eqnarray}
d_iS&=&\Delta\left(\frac{1}{T}\right)dq+\frac{1}{T}\Delta Ydx
     \nonumber\\
&&+\sum_j\frac{1}{T}\Delta\mu_{j}^{\prime}dN_j+\Delta \left(\frac{p}{T}\right)dV.\;\;\;\;
\end{eqnarray}

In the second and the third terms,  $\Delta Ydx$ and $\Delta\mu_{j}^{\prime}dN_j$ denote the heat  productions, $T$ is the temperature of the energy conversion,  so $\Delta$ does not include $1/T$.

Such that, we obtain the total differential expression for the total entropy production.

Now we continue to discuss the second law.

We consider the independent processes first, the independent process denotes that there is only one gradient of thermodynamic force driving the spontaneous process.

(1). The first type of the independent process is the heat transfer, the heat energy $dq$ is transferred between different\\ temperatures, the driving force of the heat transfer is the gradient of the temperature $\Delta T$, and the direction of the heat flux is decided by the gradient of the temperature $\Delta T$, thus, the heat energy can only be transferred from higher temperature $T_h$ to lower temperature $T_c$.
\begin{eqnarray}
dq(T_h)\rightarrow dq(T_c).
 \nonumber
\end{eqnarray}

We have the  calorimetric entropy production
\begin{equation}
d_iS=\left(\frac{1}{T_c}-\frac{1}{T_h}\right)dq=\Delta\left(\frac{1}{T}\right)dq\ge0.
\end{equation}

The heat transfer obeys such a principle: the direction of the spontaneous process is decided by the gradient $\Delta T$, the temperature gradient $\Delta T$ spontaneously tend to zero, the calorimetric entropy production describes the dissipation of the driving force $\Delta T$. In an independent process, the dissipation of the driving force $\Delta T$ or the result $\Delta(1/T)$  cannot be spontaneous recovery completely.

Eq.(74) is actually a proof for the physical meaning of the calorimetric entropy.

$dq$ is a heat element, $T$ is the intensity distribution of $dq$, named as the distribution of the energy level for $dq$.

The integral is the sum of all $dq/T$
\begin{equation}
S_q=\int_{(0]}^q\frac{dq}{T}.
\end{equation}

By Eq.(74), when $q$ is fixed and unchanging, the bigger $S_q$ is, the more evenly-distributed for the distributions of the energy levels of all $dq$,
when a heat element $dq$ transfer from a high energy level $T_h$ to a low energy level $T_c$, more of the lower energy levels are occupied, this deterministic explanation runs through down to the micro  levels.

This also means that thermodynamics itself can explain the physical meaning of the entropy.

(2). The second type is the independent process driven only by the pressure gradient,  the direction of the process is decided by the direction of the gradient $\Delta p\geq0$.

Suppose a system changes from the state $A$ to the state $B$  without other changes,
since the heat partial pressure $p$  points always to the normal direction of the given system's surface,
so the volume change $dV$ driven by the pressure $p$ always be positive definite,
the given system will generate a spontaneous diffusion process.

Since $p/T$ is a positive definite function, and in integral interval [A,B], the integral variable $dV\geq 0$ , we have
\begin{equation}
d_iS=\left(\frac{p_B}{T_B}-\frac{p_A}{T_A}\right)dV=\Delta \left(\frac{p}{T}\right)dV\ge0.
\end{equation}

At the macro level, Eq.(76) shows the entropy production of the diffusion process caused by the density gradient.

The spontaneous diffusion process driven by the pressure $p$ obeys such a principle: the direction of the process is decided by the pressure gradient $\Delta p$, the gradient $\Delta p$ spontaneously tends to zero. In an independent process, the dissipation of the the driving force $\Delta p$ or the result $\Delta(p/T)$ cannot be spontaneous recovery completely.

Next, we discuss the coupling processes.

In general, a thermodynamic coupling process involves two or more gradients of thermodynamic driving forces, and involves the two basic types that: (1), the free energies are converted into the heat energy. (2), the heat energy is converted into the free energies. These two coupling processes will be discussed below, respectively.

(1). The first type of the coupling process is the energy conversion, the free energies are converted into the heat energy directly without an adequate compensation.
\begin{eqnarray}
dA+dG\rightarrow dq.
 \nonumber
\end{eqnarray}

The driving forces of the processes are $\Delta Y$ and/or $\Delta\mu_{j}^{\prime}$, the conversion processes obey the law of the conservation of energy, so $d_iq$ cannot be less than zero, thus we get
\begin{eqnarray}
d_iq=\Delta Ydx+\sum_j\Delta\mu_{j}^{\prime}dN_j\geq0.
\end{eqnarray}

The calorimetric entropy production
\begin{eqnarray}
d_iS_q\!=\frac{d_iq}{T}=\!\frac{1}{T}\Delta Ydx\!+\sum_j\frac{1}{T}\Delta\mu_{j}^{\prime}dN_j\geq0.
\end{eqnarray}
where $T$ is the temperature of the heat conversion.

The production of the calorimetric entropy $d_iq$ will be accompanied with a change of the compensating potential,
the compensation relationship describe the reversible parts, by Eq.(21), it follows that
\begin{eqnarray}
d_iq=-pdV.
\end{eqnarray}

Since $p$ is positive definite, where $dV\leq 0$.

The free energy converting into the heat energy will drive an increase of the compensating potential $pdV\leq0$.

In an irreversible process,
the spontaneous dissipation of the compensating potential is always positive definite, so by Eq.(22), it follows that
\begin{eqnarray}
pdV\geq-d_iq.
\end{eqnarray}

Consider the given system changes from the state A to the state B, according to Eq.(78), Eq.(79) and Eq.(80), we obtain the total entropy  production
\begin{eqnarray}
d_iS\!=\!\frac{1}{T}\Delta Ydx\!+\sum_j\frac{1}{T}\Delta\mu_{j}^{\prime}dN_j+\Delta \bigg(\frac{p}{T}\bigg)dV\geq0.\nonumber\\
\end{eqnarray}

$d_iS$ represents the free energy dissipations in the energy conversion processes.
The energy conversion processes are derived from such a principle: the gradients $\Delta Y$ and $\Delta \mu_j^{\prime}$ spontaneously tend to zero.

(2), The second type of the coupling process is also the energy conversion, the heat energy  is converted into the free energies.
\begin{eqnarray}
dq\rightarrow dA+dG.
 \nonumber
\end{eqnarray}

The driving forces of the processes is $\Delta p$, the conversion processes obey the law of the conservation of energy, so $d_iq$ is less than zero, we get
\begin{eqnarray}
d_iq=\Delta Ydx+\sum_j\Delta\mu_{j}^{\prime}dN_j\leq0.
\end{eqnarray}

The calorimetric entropy production
\begin{eqnarray}
d_iS_q\!=\frac{d_iq}{T}=\!\frac{1}{T}\Delta Ydx\!+\sum_j\frac{1}{T}\Delta\mu_{j}^{\prime}dN_j\leq0.
\end{eqnarray}

When the heat energy is converted into the free energies, there must be some compensation,
the driving forces of the processes come from the compensating potential,
the compensation relationship describe the reversible parts.
\begin{eqnarray}
d_iq=-pdV.
\end{eqnarray}

In an irreversible process, there are some spontaneous dissipations of the compensating potential, so we have
\begin{eqnarray}
pdV\geq -d_iq.
\end{eqnarray}

We obtain
\begin{eqnarray}
d_iS\!=\!\frac{1}{T}\Delta Ydx\!+\sum_j\frac{1}{T}\Delta\mu_{j}^{\prime}dN_j+\Delta\bigg(\frac{p}{T}\bigg)dV\geq0.\nonumber\\
\end{eqnarray}

Eq.(86) represents the dissipations of the compensating potential  in the energy conversion processes. The dissipations of the compensating potential are derived from such a principle: the gradients $\Delta p$ spontaneously tend to zero.

In the compensation equation for reversible processes, Eq.(79) and Eq.(84) show the  two different situations: in Eq.(79), we have the heat production $d_iq\geq 0$, meanwhile the increment of the compensating potential $pdV\!\leq 0$; in Eq.(84), we have the heat production  $d_iq\leq 0$, meanwhile the increment of the compensating potential $pdV\!\geq 0$.

By Eq.(79) and Eq.(84), in Eq.(81) and Eq(86), we can see the compensating relationship for the reversible parts.
\begin{eqnarray}
d_iS_{rev}=\frac{1}{T}d_iq+\frac{p}{T}dV=0.
\end{eqnarray}
where, the subscript $rev$ denotes a reversible process.

By definitions, Eq.(87) can also be written as
\begin{eqnarray}
d_iS_{rev}=d_iS_q+d_iS_v=0.
\vspace{0.2cm}
\end{eqnarray}
where, $d_iS_q$ is the calorimetric entropy production, and $d_iS_v$ is the configurational entropy production.

In a reversible coupling process, by Eq.(79) and Eq.(81), when the free energies $A$ and/or $G$  convert into the heat energy $q$, we know that the calorimetric entropy production $d_iS_q\geq0$, but the configurational entropy production $d_iS_v\leq0$,
by Eq.(84) and Eq.(86), when the heat energy $q$ is converted into the free energies $A$ and/or $G$, we have the calorimetric entropy production $d_iS_q\leq0$,
but  the configurational entropy production $d_iS_v\geq0$.

This is an important conclusion: in a coupling process,
we always have the total entropy production $d_iS\geq 0$, but the partial differential of the total entropy production can be less than zero.
The second law itself has already contained the internal mechanism of the order production.

Under more general situations, irreversible processes usually are the mixed behaviors of the above processes that we discussed respectively.
Combine with Eqs.(74), (76), (81) and (86), we obtain
\begin{eqnarray}
d_iS&=&\Delta \left(\frac{1}{T}\right)dq+\frac{1}{T}\Delta Ydx\nonumber\\
&&{}+
\sum_j\frac{1}{T}\Delta \mu_{j}^{\prime}dN_j+\Delta \left(\frac{p}{T}\right)dV\ge0.
\end{eqnarray}

There are three positive definite sources of the entropy productions:
(1), Heat transfer from higher temperature to lower temperature;
(2), The spontaneous dissipation of the compensating potential; (3). Free energies  convert into heat energy without an adequate compensation.

In an independent process, the direction of the gradient of the thermodynamic driving force decides the direction of the spontaneous process.

In an irreversible process, the total entropy production is always positive definite,
we have $d_iS\!\ge0$, but the partial differentials of which can be less than zero,
the second law itself has already contained the mechanism of evolution, it implies that order and evolution can arise spontaneously.

The second law of thermodynamics elucidates such a fundamental principle of nature:

Thermodynamic processes are driven by the gradients of thermodynamic forces, all of the gradients of the thermodynamic forces spontaneously tend to zero,
as a result, the total entropy production $d_iS\!\ge0$ indicates the dissipations of the thermodynamic forces,
and according to Eq.(89), the dissipations of the gradients of the thermodynamic driving forces cannot be spontaneous recovery completely.

In an independent process, the direction of a spontaneous process is decided by the direction of the gradient of the thermodynamic force,
on the other hand, in a coupling process, the dissipations of one thermodynamic forces can drive the gradient production of another thermodynamic force,
therefore, the partial differential of the total entropy production can be less than zero, the second law itself has already contained the mechanism of evolution.

\section{\label{sec:level1}a groundbreaking overall progress\protect
}
Now, let's review on the fundamental problems of the traditional thermodynamics.

1, We didn't know the intensity distributions indicated  by the temperature $T$ and the pressure $p$ respectively should
be anchored to which extensive quantity.

2, Meanwhile, we didn't know which state quantity that the process quantity $\delta Q$ is anchored with.

3, The temperature $T$ of a system and the pressure $p$ of a system are only applicable to equilibrium state,
therefore, the evolutionary processes from an initial to an final states must be assumed to be a series of quasi-static processes.

4, In the traditional thermodynamics, there has neither a state quantity of heat, nor dynamics.

5, We didn't know which state quantity that $\delta Q/T$ can be anchored to, the traditional  thermodynamics itself cannot explain what the physical meaning of the entropy is.

6, The total entropy production
$d_iS\geq0$ is only a result rather than the internal mechanism.

and more \ldots

In the absence of a fundamental physical quantity \ldots

This absent concept is an essential concept in the whole theoretical framework of thermodynamics, it is no doubt, if no this concept,
thermodynamics will always face an unsolvable theoretical predicament, in fact, this predicament has being continued for more than a century.

What we are seeing is a groundbreaking overall progress, the predicament of equilibrium state has been  completely broken through, almost all of the
long-standing problems in traditional thermodynamics will be finally clarified, what we are seeing is a hopeful
forefront: a perfectly theoretical framework of thermodynamics, and what we can also see is a series of new developments in different fields.

In a new theoretical framework, each of the intensive quantities has been  anchored to a corresponding extensive quantity, and meanwhile, each of the process quantities has a corresponding state quantity as its anchoring, we don't need to be bothered anymore by those concepts in classical thermodynamics that  always can hardly  be explained.

In a new theoretical framework, the narrative paradigm of thermodynamics has  been changed significantly, and has been replaced by a new narrative paradigm of dynamics:
 thermodynamic processes are driven by the gradients of the different thermodynamic forces, the relationship between the state and changes
of a thermodynamic system and the driving forces are the main themes of thermodynamics.

In physical chemistry, when chemical thermodynamics itself unfold in the theoretical form of dynamics, there is no longer a distinction between chemical thermodynamics and chemical dynamics\! /\! kinetics.

In physics, the relationship between thermodynamics and statistical physics is facing a revalidation, because now thermodynamics itself can already provide a deterministic  explanation of the physical contents of the entropy, which is different from the prevalent probabilistic explanation, and an important result is, this deterministic explanation runs through down to the microscopic levels.

In evolutionary physics\cite{Prigogine1980}, the order production and the dissipation both follow a fundamental  law: the second law of thermodynamics, this is a grand theme that penetrates through all levels: from the tiniest physical reality levels all the way to astrophysics, involves our view of nature.

and more \ldots

In a new theoretical framework \ldots

We are witnessing a profound change in the history of thermodynamics, we are understanding what we could not understand in the past, and we are reestablishing a grand scientific theoretical  framework  that hasn't been completed for over a century, these kinds of the situations and the new opportunities rarely happen in the history of science.

\end{document}